# Effect of Demonetisation of Indian high denomination currencies on Indian Stock Market and its relationship with Foreign Exchange Rate

By
Dr.N.Suresh[1], & Bharathi N.R[2]
1-Professor, 2-Faculty
1. Faculty of Management and Commerce, Ramaiah University of Applied Sciences, Bangalore 560 054
2. BSE Group of Institutions, Bangalore

**Abstract:**

This paper examines the impact of foreign exchange rate i.e.US Dollar to Indian Rupee (USD/INR) on Indian Stock Market Index (Nifty 50) during demonetisation of high denomination Indian Currencies. Daily rate of return of foreign exchange rate (USD/INR) and Indian Stock Market Index (Nifty 50) were considered for the study. 'Dummy' variable was used to measure the effect of demonetisation during Nov/Dec 2016. The period of study was restricted to 243 days from 1st April 2016 to 31st March 2017. Study reveals that there was an upward trend observed in Indian stock market and Indian currency was strengthened with the decrease in foreign exchange rate (USD/INR).

Key words: Exchange rate, Indian stock market, ARCH GARCH

## 1. Introduction:

November 8th 2016 when entire global was anxiously looking for presidential election in US, Indian Prime Minister Narendra Modi announced that the high denomination notes of ₹.500 and ₹.1,000 would cease to be legal. 50 days period was given to Indian population to exchange their old 500 and 1000 rupees notes with new currencies by depositing them into bank. It was a drastic monetary step in which a currency units states as a legal tender is declares invalid. This is done whenever there is a change of national currency, replacing the old unit with a new one. It aimed to wash-down the stock of black money supply and to remove the fake note out of economy (Law, W. Y., 2013). After demonetization of Indian currency, rupee has become stronger against good number of currencies including US Dollar ($), Euro, Australian Dollar, Singapore Dollar, Japanese Yen, British Pound, and had become weaker against few currencies viz., Russian Ruble and South African Rand. It was also observed that, Indian stock market also reacted positively with demonetisation.

## 2. Theoretical Background:

Demonetisation had effected the cash transaction as there were shortages of lower denomination currencies like ₹.50 & ₹.100. During demonetization payment became big problem. Many micro/tiny businesses were shut down, and many days there was no payment for working class, especially daily workers like painter, carpenter, plumber, electrician, this resulted in unemployment. Totally it disturbed the ordinary people and livelihood of weaker sections of society. According to Nikkei India, Manufacturing purchasing manager's index (PMI) fell to from 52.3 to 49.6 in indicating a downturn in the overall manufacturing activity. Cash crunch has resulted in lower purchasing activity in turn leading to lower manufacturing output. Cash delivery business came to standstill. Real estate business and other sectors depended on cash transaction suffered loss. This reveals that all the sectors of Indian stock market affected because of demonetisation. Hence, the study on exploring the interaction between Indian stock Market and foreign exchange rate found optimal (Murthy, U., Anthony, P., & Vighnesvaran, R., 2016)

One of the significant indicators of vigorous fluctuations in stock prices is volatility of stock market. (Selvam, 2011). Considerate volatility in stock markets is vital for the assessment of investment and leverage decisions as it has substantial impact on risk-averse investors (Balasubramanyam, 2003). Exchange Rate is one of the most important factor for any economic growth of any country, it has direct effect on international trade. Exchange rate can be best comprehended as just a benchmark for a country's cash supply. Due to the ban of ₹ 500 and ₹ 1000 denominator currency by the Indian government, has impacted the exports in sectors like agriculture due to high number of cash transactions. This ban greatly impacted the growth of export – import EXIM trade. (Maersk, 2017). Financial economists, academicians and policy makers had considered movement of Exchange rate, and fluctuations in stock markets as an important subject of analysis (Zakaria, 2012),(Musa, S.,2014). Subsequently, a good number of models were established in the literacy of finance for the investigation and the measurement of volatility (Elsherif, M. A., 2016). Autoregressive conditional heteroscedastic (ARCH) model by Engle (1982) and generalized (GARCH) model by Bollerslev (1986) and Taylor (1986) were developed to forecast the volatility.

### 3. Objective of the study:

Objective is to develop general to specific GARCH (1, 1) Model to check whether demonetisation of high value Indian currency notes has an impact on Indian Stock Market return. (Bala, D. A., & Asemota, J. O., 2013).

Following are the steps involved. (Adesina, K. S., 2013)

- Analyse the trend of Indian stock market, foreign exchange rate (USD/INR) and their daily rate of return. (Akhtar, S., Ansari, V. A., & Ansari, S. A. 2017)
- Find out the order of integration of the variables by Unit root Test.
- Run regression model with Stock Market return as dependent variable and return on foreign exchange rate as exogenous variable to generate residuals.

- Perform ARCH test for residuals to check heteroscedasticity for running ARCH family model.
- Developing ARCH family model with three types of distribution in variance equation viz., Normal Gaussian distribution, Generalised Error distribution and student's t with fixed degree of freedom with dummy variable to check the effect of demonetisation effect. (Epaphra, M.,(2017), (Mathur, S., Chotia, V., & Rao, N. V. M., 2016).
- Diagnostic checking of the model with Stock Market return as dependent variable and return on foreign Exchange rates as exogenous variable along with dummy variable.
- Suggest the desirable model in which residuals does not have serial correlation, no ARCH effect and non-normality. (Srinivasan, P. 2011).

## 4. Data, Results and Discussion

### 4.1 Analysis of Indian Stock Market and Exchange rates and its return:

We have used two variables such as Indian Stock Market Index (Nifty 50) and United States Dollar (USD) in generic model (Murari, K., 2015), (Oberholzer, N., & Venter, P., 2015). It is a time series model from 1/04/2016 to 31/03/2017. The descriptive statics representing the daily data and its retutn are shown in Table 1. (Alom, F., 2016)

**Table 1: Descriptive Statistics of the variable**

| Table 1: Descriptive Statistics of the variable | | | | |
|---|---|---|---|---|
| Variable | Nifty | USD/INR | Return on NIFTY | Return on USD/INR |
| OBS | 248 | 248 | 243 | 243 |
| Mean | 8421.20 | 67.04 | 0.0725 | -0.0077 |
| SE MEAN | 24.80 | 0.05 | 0.0506 | 0.0187 |
| StDev | 390.00 | 0.73 | 0.7885 | 0.2917 |
| Minimum | 7546.40 | 64.84 | -2.6913 | -1.1574 |
| Median | 8502.90 | 66.93 | 0.0694 | 0.0000 |
| Maximum | 9173.80 | 68.81 | 2.4010 | 0.9569 |
| Skewness | -0.17 | -0.31 | -0.1400 | -0.1900 |
| Kurtosis | -0.85 | 1.02 | 1.1100 | 1.3200 |

During the study period it was found that Exchange rate of Dollar has reached Maximum of 68.81 and minimum of 64.84 with a mean of 67.04. It was also observed that Nifty has reached its peak 9173.80 and had reached as low as 7546.40.

During the study period it was also found that return on Exchange rate of Dollar has reached Maximum of 0.95 and minimum of -1.15 with a mean of 0.0. It was also observed that return on Nifty has reached maximum of 2.40 and with minimum of -2.67 with mean of 0.07

The trend line of data and return for the foreign exchange rate are shown in Fig.1. It shows that the there is a downward trend for both exchange rates at the end of the study period. The return on foreign exchange rate found stationary (Sekmen, T., & Hatipoglu, M., 2016)

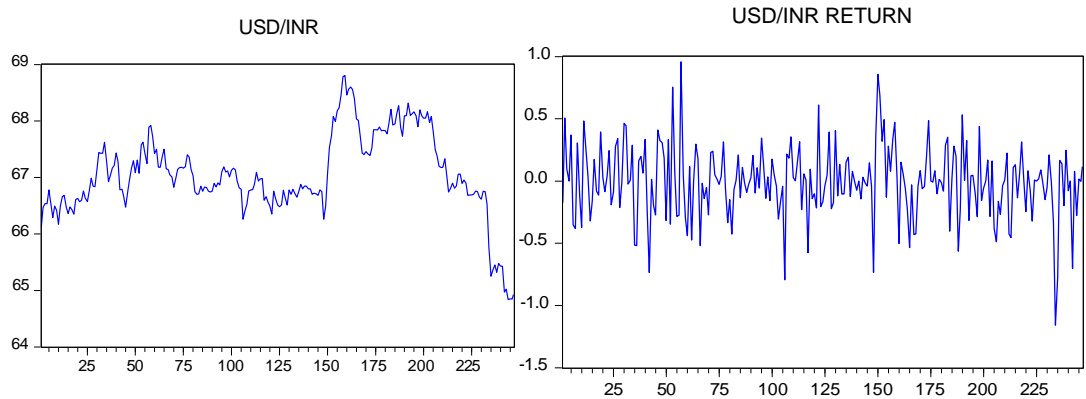

Fig.1: Daily Data and Daily Reurn data of US Dollar (USD)

The trend line of data and return for the Indian Stock market (Nifty) are shown in Fig.2. (Angabini, A., & Wasiuzzaman, S., 2011). It depicts that the trend was almost stationary from 150 to 185 days and then there is continuous upward trend in the Indian Market. Return on Stock market was found stationery. (Kamble, G., & Honrao, P.,2014).

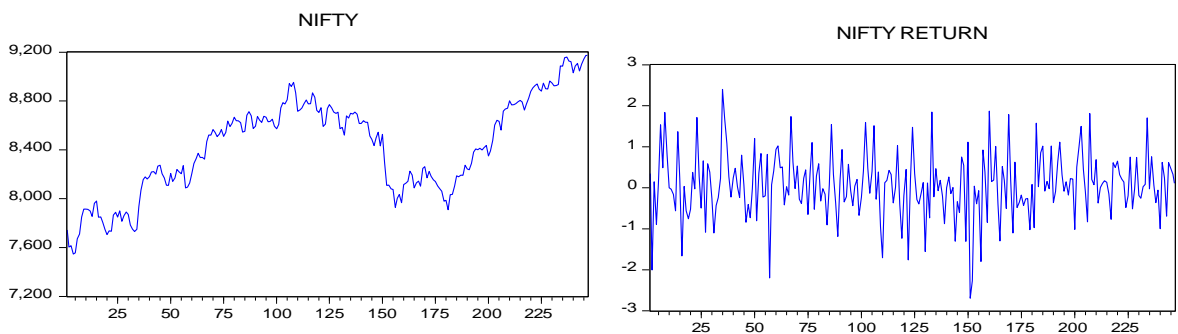

**Fig.2: Daily Data and Return of Nifty**

From the above analysis it was found that, during 150-185 days, the Indian Stock Market had shown sensitiveness indicating that 'Demonetisation of Higher denomination Currency' has made a greater impact on stock market. In order to identify and measure this effect, we has used another variable named as 'Dummy'.

**Model:**

We have used three variables such as Indian Stock Market Index (Nifty 50), United States Dollar (USD) and Dummy variable in this model. Here our data for dummy is from 9/11/2016 to 31/12/2017. Here we have used dummy variable (Dummy) to capture the effect of Demonetisation on Indian Stock market. Here, Dummy = 0, prior to demonetisation period 1/4/2016 to 8/11/2016) and after demonetisation period (1/1/2017 to 31/03/2017), while Dummy = 1 during demonetisation period (9/11/2017 to 31/12/2017).

If the Dummy variable is found significant and positive, demonetisation of high denomination Indian Currency has a positive impact on Nifty along with the foreign exchange rate of two currencies. (Amatyakul, D., & Chintrakarn, P., 2012).

**4.2 Unit Root Test (Stationary Test):**

We expect that time series data should **stationary** to run Auto-regressive models at levels. Hence, Stationery test for different data series were conducted through Augmented Dickey Fuller Test (ADF).

The results of ADF shown in Table 2.

| Table 2: Unit Root Test of Daily Return of Nifty and Foreign Exchange Rate | | | | | |
|---|---|---|---|---|---|
| | | NIFTY | | USD/INR | |
| Augmented Dickey-Fuller Test Statistics | | t-statistics | Prob. | t-statistics | Prob. |
| | | -15.2807 | 0.0000 | -14.4027 | 0.0000 |
| Test Critical Values | 1% Level | -3.4568 | | -3.45684 | |
| | 5% Level | -2.8730 | | -2.87309 | |
| | 10% Level | -2.5730 | | -2.57300 | |
| *Source: Calculated from the data taken from NSE and Quandl.com.com website for a selected period using E-Views Software* | | | | | |

The ADF statistic value for Nifty is -15.2807, USD/INR is -14.4027 and the associated one-sided p-value (for a test with 243 observations) is 0.00 for all the variables. Test Critical values at the 1%, 5% and 10% levels are shown in table 2. It was found that, the t-statistics is less than the critical values meaning that variables does not have unit root and is desirable at level.

**4.3 Regression Model:**

The regression estimation with the variables viz., retun on Indian Stock Market (Nifty) and US Dollar (USD_IND) are shown in equation (1).

Mean equation:
$$NIFTY = C_1 + C_2 * USD\_INR + e \text{ ----- (1)}$$

Here,

Nifty = Return on stock Market return, $C_1$= constant,

USD_INR = Return on Exchange rate of US Dollar,

e = residual. Here we have taken return on daily data of 243 days

The result of estimated regression equation is shown in Table 3.

The test of significance was done by T-statistics.

T-statistic, is used to examine the hypothesis that the estimated coefficient is equal to zero or not. P-value is used to interpret the t-statistic. When the P-value is less than or equal to 5% (0.05), then the independent variable is significant to influence the dependent variable. F-statistics reported in the regression output is from a test of the hypothesis that all of the slope of the coefficients in a regression are zero. F-Test is a joint test, which can be highly significant, even if t-statistics are insignificant for the variables. Means Independent variables should be jointly significant to influence dependent variable. If the p-value is less than the significant level of testing, reject the null hypothesis that all slope coefficients are equal to zero.

| Table 3: Results of Estimated Regression Model ||||||
|---|---|---|---|---|---|
| NIFTY = $C_1$ + $C_2$*USD_INR + e------ (1) ||||||
| Variable | Name of the Variable | Estimated Coefficient | Std Error of Coefficient | T-Statistic | Probability |
| Constant | C1 | 0.065626 | 0.047851 | 1.371449 | 0.1715 |
| USD_INR | C2 | -0.893770* | 0.164304 | -5.439741 | 0.0000 |
| R-squared | | 0.109356 | Akaike info criterion || 2.259136 |
| F-statistic | | 29.59078 | Schwarz criterion || 2.287885 |
| Prob(F-statistic) | | 0.000000* | Durbin-Watson statistic || 2.066633 |
| *Significant at 1% ||||||

### 4.4 ARCH test:

**In order to fit ARCH/GARCH Model we should run the regression model to see that the residuals should be stationery. The graphical representation of residual data are shown in Fig 3**

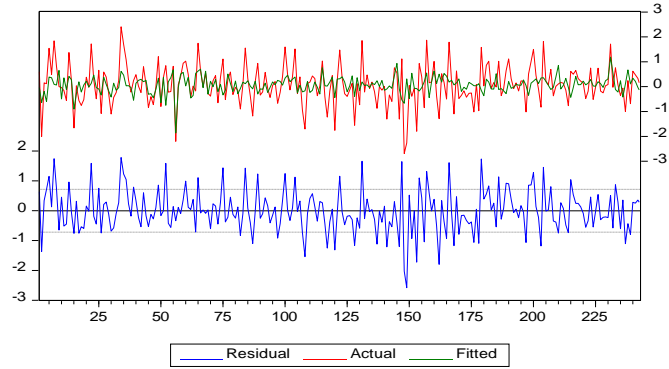

Fig 3. Residual data graph

It is found that periods of low volatility followed by periods of low volatility and periods of high volatility is followed by periods of high volatility for a continued period indicating volatility clustering. When this is observed for the residuals, then we can justify to run ARCH family model. This can further investigated by adopting ARCH test. (Abdalla, S. Z. S., 2012), (Akhtar, S., Ansari, V. A., & Ansari, S. A. 2017).

Heteroskedasticity LM test will be conducted with the following Hypothesis.

 Null: There is no ARCH affect

 Alternative; There is ARCH effect
The results are shown below in table 4.

| Table 4. Heteroskedasticity Test | | | |
|---|---|---|---|
| F-Statistics | 5.3231 | Probability F(1.240) | 0.0219 |
| Obs.* R-squared | 5.2510 | Prob. Chi-Square | 0.0219 |

 It is found that Probability of Chi-Square is 0.0219 which is less than 5 percent meaning that we can reject null hypothesis and we can accept alternative hypothesis. So, we can run the ARCH/GARCH Model.

### 4.5 Development of GARCH (1, 1) Model:

The development of General GARCH (1.1) Model consists of two equations viz., Mean Equation and Variance Equation.

Mean Equation is represented by

NIFTY = $C_1$ + $C_2$*USD_INR + e ------(1)

and Variance equation is represented by

GARCH = $C_3$ + $C_4$*(RESID (-1))$^2$ + $C_5$*GARCH (-1) + $C_6$* DUMMY + e ------ (2

GARCH = Residual Variance (error term) derived from the mean equation (1). It is also known as present/current day's variance or volatility of Stock Market Return (NIFTY). (Olweny, T., & Omondi, K., 2011).

(RESID (-1))$^2$ = Previous periods' residual square obtained from equation (1) .and is known as Lag /previous day's return information regarding volatility. It is known as ARCH Term

GARCH (-1) = Lag/Previous day's variance residual or volatility of Stock Market Return (NIFTY). Term is known as GARCH

DUMMY = Variable to represent the effect of demonetisation

**4.6 Results Discussion of GARCH (1, 1) Model:**

The study has been done with Nifty as dependent variable, USD/INR as exogenous variable in mean equation and DUMMY as exogenous variable in variance equation (Bucevska,V, 2013). Three types of distribution viz., Normal Gaussian, Generalised Error and Student's t distribution with were used in this study. The results were shown in table 5 to table 7.

**4.6.1 Normal Gaussian distribution:**

Here, Coefficients of Return on exchange rate USD/INR (-0.7500) is found significant meaning that the return on exchange rate can influence the volatility of Nifty. GARCH term is found to be significant with the coefficient of -0.55574. It means that lag period/ previous day Nifty's return volatility ($C_5$ in equation 2) can influence today's Nifty's volatility of Return. In addition DUMMY variable also found highly significant at 1% with 1.2325. It means that Volatility of Nifty Return is influenced by GARCH, DUMMY factor and the exchange rates.

| Table 5: Results of GARCH (1,1) Model |||||||
|---|---|---|---|---|---|---|
| NIFTY = $C_1$ + $C_2$*USD_INR + e------ (1) |||||||
| GARCH = $C_3$ + $C_4$*(RESID (-1))$^2$ + $C_5$*GARCH (-1) + $C_6$* DUMMY + e--------(2) |||||||
| Variable | Name of the Variable | Estimated Coefficient | Std Error of Coefficient | Z-Statistic | Probability ||
| Constant | C1 | 0.1301 | 0.0444 | 2.9317 | 0.0034 ||
| USD_INR | C2 | -0.7500* | 0.1312 | -5.7156 | 0.0000 ||
| Variance Equation with Normal Gaussian Distribution |||||||
| Constant | C3 | 0.7221* | 0.1194 | 6.0474 | 0.0000 ||
| (RESID (-1))$^2$ | C4 | -0.0555 | 0.0344 | -1.6142 | 0.1065 ||
| GARCH (-1) | C5 | -0.5574* | 0.1895 | -2.9402 | 0.0033 ||
| DUMMY | C6 | 1.2325* | 0.2215 | 5.5645 | 0.0000 ||
| R-squared || 0.109356 | Akaike info criterion || 2.259136 ||
| Durbin-Watson statistic || 2.066633 | Schwarz criterion || 2.287885 ||
| *Significant at 1% |||||||

### 4.6.2 Generalised Error Distribution Assumption

| Table 6: Results of GARCH (1,1) Model |||||||
|---|---|---|---|---|---|---|
| NIFTY = $C_1$ + $C_2$*USD_INR + e------ (1) |||||||
| GARCH = $C_3$ + $C_4$*(RESID (-1))$^2$ + $C_5$*GARCH (-1) + $C_6$* DUMMY + e--------(2) |||||||
| Variable | Name of the Variable | Estimated Coefficient | Std Error of Coefficient | Z-Statistic | Probability ||
| Constant | C1 | 0.1091 | 0.0414 | 2.6320 | 0.0085 ||
| USD_INR | C2 | -0.6705* | 0.1313 | -5.1064 | 0.0000 ||
| Variance Equation with Normal Generalised Error Distribution |||||||
| Constant | C3 | 0.7030* | 0.1495 | 4.7012 | 0.0000 ||
| (RESID (-1))$^2$ | C4 | -0.0592 | 0.0357 | -1.6583 | 0.0972 ||
| GARCH (-1) | C5 | -0.5285* | 0.2266 | -2.3321 | 0.0197 ||
| DUMMY | C6 | 1.2531* | 0.3154 | 3.9723 | 0.0001 ||
| GED Parameter || 1.4108* | 0.2275 | 6.2011 | 0.0000 ||
| R-squared || 0.0997 | Akaike info criterion || 2.1775 ||
| Durbin-Watson statistic || 2.0323 | Schwarz criterion || 2.2781 ||
| *Significant at 1% |||||||

Here, Coefficients of Return on exchange rate USD/INR (-0.6705) is found significant meaning that the return on exchange rate can influence the volatility of Nifty. GARCH term is found to be significant with the coefficient of -0.5285. It means that lag period/ previous day Nifty's return volatility ($C_5$ in equation 2) can influence today's Nifty's volatility of Return. In addition DUMMY variable also found highly significant at 1% with 1.2531. It means that Volatility of Nifty Return is influenced by GARCH, DUMMY factor and the exchange rates.

### 4.6.3 Student's t Distribution:

Here, Coefficients of Return on exchange rate USD/INR (-0.6855) is found significant meaning that the return on exchange rate can influence the volatility of Nifty. GARCH term is found to be significant with the coefficient of -0.4876. It means that lag period/ previous day Nifty's return volatility ($C_5$ in equation 2) can influence today's Nifty's volatility of Return. In addition DUMMY variable also found highly significant at 1% with 1.3162. It means that Volatility of Nifty Return is influenced by GARCH, DUMMY factor and the exchange rates.

| Table 7: Results of GARCH (1,1) Model | | | | | |
|---|---|---|---|---|---|
| NIFTY = $C_1$ + $C_2$*USD_INR + e------ (1) | | | | | |
| GARCH = $C_3$ + $C_4$*(RESID (-1))$^2$ + $C_5$*GARCH (-1) + $C_6$* DUMMY + e--------(2) | | | | | |
| Variable | Name of the Variable | Estimated Coefficient | Std Error of Coefficient | Z-Statistic | Probability |
| Constant | C1 | 0.1130 | 0.04313 | 2.6204 | 0.0088 |
| USD_INR | C2 | -0.6855* | 0.1279 | -5.3590 | 0.0000 |
| Variance Equation with Students' t Distribution | | | | | |
| Constant | C3 | 0.6868* | 0.1710 | 4.0157 | 0.0001 |
| (RESID (-1))$^2$ | C4 | -0.0595 | 0.03228 | -1.8438 | 0.0652 |
| GARCH (-1) | C5 | -0.4876* | 0.2144 | -2.2738 | 0.0230 |
| DUMMY | C6 | 1.3162* | 0.5113 | 2.5793 | 0.0101 |
| T-Dist. DOF | | 7.7117 | 6.1021 | 1.2647 | 0.2059 |
| R-squared | | 0.10000 | Akaike info criterion | | 2.1909 |
| Durbin-Watson statistic | | 2.0329 | Schwarz criterion | | 2.2916 |
| *Significant at 1% | | | | | |

### 4.7 Residual Diagnostic Test for GARCH (1, 1) Model:

Residual diagnostic was conducted for GARCH (1, 1) Model to select the desirable model to measure the volatility of stock market return with the effect of demonetisation. (Ali, T. M., Mahmood, M. T., & Bashir, T. 2015).

**Testing for serial Correlation:**

Estimated equation should be examined for the evidence of serial correlation by applying Correlogram square residual (Q test) test. Following are the assumptions of the test.

- Null: There is no serial correlation in the residual or error term
- Alternative: There is serial correlation

Results are shown in table 8. Q-statistics are insignificant in all distribution with large p-values > 0.05, we accept null hypothesis indicating that there is no serial correlation in the residuals.

| Table.8. Correlogram of Squared Residual | | | | | | |
|---|---|---|---|---|---|---|
| | Normal Gaussian Distribution | | Generalised Error Distribution | | Student's t Distribution | |
| Lag | Q-stats | Probability | Q-stats | Probability | Q-stats | Probability |
| 1 | 0.1302 | 0.718 | 0.1831 | 0.669 | 0.2261 | 0.634 |
| 2 | 1.8980 | 0.387 | 2.1954 | 0.334 | 2.4098 | 0.300 |
| 3 | 3.0285 | 0.387 | 4.4167 | 0.220 | 4.8498 | 0.183 |
| 4 | 5.5086 | 0.239 | 8.3494 | 0.080 | 8.7262 | 0.068 |
| 5 | 7.1555 | 0.209 | 8.8900 | 0.114 | 9.2951 | 0.098 |

2. **Heteroskedasticity LM Test:**

This ARCH test is to check whether residuals are auto correlated. Following are hypothesis.
- Null: There is No Arch Effect
- Alternative: There is ARCH effect

Results are shown in table 9.

| Table 9. ARCH Test | | | |
|---|---|---|---|
| | Normal Gaussian Distribution | Generalised Error Distribution | Student's t Distribution |
| F-Statistics | 0.5858 | 0.9195 | 0.9414 |
| Probability F(1.240) | 0.4448* | 0.3386* | 0.3329* |
| Obs.* R-squared | 0.5893 | 0.9236 | 0.9455 |
| Prob. Chi-Square | 0.4427* | 0.3365* | 0.3308* |

It is found that probability of Chi-Square is more than 5 percent meaning that we can accept null hypothesis which says that there is no ARCH effect.

1. **Normality Test:**

Residual Diagnostics/Histogram–Normality Test describes shows descriptive statistics and a histogram of the residuals. Jarque-Bera statistic is the component to diagnose whether the residuals are normally distributed or not. If the standardized residuals are normally distributed, the Jarque-Bera statistic should not be significant. Results are shown in table 10.

- Null: Residuals are normally distributed
- Alternative: Residuals are normal

| Table 10. Normality Test | | | |
|---|---|---|---|
| | Normal Gaussian Distribution | Generalised Error Distribution | Student's Distribution |
| | | | |
| Jarque-Bera Statistics | 2.5962 | 2.6088 | 2.5963 |
| Probability | 0.2730* | 0.2713 | 0.2731 |
| Skewness | 0.1229 | 0.0785 | 0.0932 |
| Kurtosis | 3.4427 | 3.4826 | 3.4706 |

The probability of Jaqua-Bera is less than 5%. Null Hypothesis is rejected for all distribution. It indicates that residuals are normal, which is desirable.

**4.8 Model Selection:**

After residual diagnostic check, it was found that GARCH (1, 1) Model with all the distribution found satisfactory with Q-Test for serial correlation, ARCH-Test for Heteroskedasticity and Normality test with Jarque-Bera statistics. Hence, all the models are accepted. Hence, desirable model will be Generalised Error Distribution with lowest AIC value of 2.1775.

**5. Conclusions:**

Modeling and forecasting the volatility of return on stock market has become an important field of empirical research in finance. This is because volatility is considered as an important concept in many economic and financial applications. This paper attempts to explore the interaction between Exchange rates and Indian Stock Market during demonetisation of Indian Currency. The volatility of the Indian Stock Market returns have been modeled by using Generalized Autoregressive Conditional Heteroscedastic (GARCH) models that captures most common stylized facts about exchange returns such as volatility clustering and leverage effect, following three residual distributions viz., normal, Student's t-distribution and Generalised Error distribution. The model is used for capturing the symmetry effect. The study used the USD/INR exchange rates data for the period 1$^{ST}$ April 2016 to 31$^{st}$ March 2017as exogenous variable and dummy variable was used to measure the impact of demonetisation. On the basis of observed results, the following are the conclusions.

Daily returns are charecterised by these models with strong confirmation. The exchange rate USD/INR showed a significant movement from normality and there exists a conditional heteroscedasticity in the residuals series. (Tripathy, S., & Rahman, A., 2013). Descriptive statistics for the USD/INR exchange rates shows existence of negative skewness and low kurtosis.

The results of the ARCH-LM test conducted point out significant presence of ARCH effect in the residuals as well as volatility clustering effect. (Mumo, M. P., 2017) All models were found satisfactory in all the diagnostic tests and Generalised Error Distribution model is desirable to select as best model with lowest Akaike Information Criteria. (Gokbulut, R. I., & Pekkaya, M., 2014)

**References**:

1. Abdalla, S. Z. S. (2012). Modelling exchange rate volatility using GARCH models: Empirical evidence from Arab countries. *International Journal of Economics and Finance*, *4*(3), 216.

2. Adesina, K. S. (2013). Modelling Stock Market Return Volatility: GARCH Evidence from Nigerian Stock Exchange. *International Journal of Financial Management*, *3*(3), 37.

3. Ali, T. M., Mahmood, M. T., & Bashir, T. (2015). Impact of Interest Rate, Inflation and Money Supply on Exchange Rate Volatility in Pakistan. *World Applied Sciences Journal*, *33*(4), 620-630.

4. Alom, F. (2016). A Note on the Asymmetry and Persistency of Shocks in Malaysian Exchange Rate Volatility. *Malaysian Journal of Economic Studies*, *53*(2), 227-238.

5. Amatyakul, D., & Chintrakarn, P. (2012). Anomalies: Evidence from Emerging Markets. *Journal of Applied Sciences*, *12*(8), 761-767.

6. Angabini, A., & Wasiuzzaman, S. (2011). GARCH Models and the Financial Crisis-A Study of the Malaysian. *The International. Journal of Applied Economics and Finance*, *5*(3), 226-236.

7. Bala, D. A., & Asemota, J. O. (2013). Exchange-rates volatility in Nigeria: Application of GARCH models with exogenous break. *CBN Journal of Applied Statistics*, *4*(1), 89-116.

8. Bucevska,V. (2013). An empirical evaluation of GARCH models in value-at-risk estimation: Evidence from the Macedonian stock exchange. *Business Systems Research*, *4*(1), 49-64.

9. Elsherif, M. A. (2016). Exchange Rate Volatility and Central Bank Actions in Egypt: Generalized Autoregressive Conditional Heteroscedasticity Analysis. *International Journal of Economics and Financial Issues*, *6*(3).

10. Epaphra, M. (2017). Modeling exchange rate volatility: application of the GARCH and EGARCH models. *Journal of Mathematical Finance*, *7*(01), 121.

11. Gokbulut, R. I., & Pekkaya, M. (2014). Estimating and forecasting volatility of financial markets using asymmetric GARCH models: An application on Turkish financial markets. *International Journal of Economics and Finance*, *6*(4), 23.

12. Kamble, G., & Honrao, P. (2014). Time-series Analysis of Exchange Rate Volatility of Indian Rupee/US Dollar-An Empirical Investigation. *Journal of International Economics*, *5*(2), 17.

13. LAW, W. Y. (2013). *Asymmetric and cross-sectional effects of inflation on Malaysian stock returns under varying monetary conditions* (Doctoral dissertation, UTAR).

14. Mathur, S., Chotia, V., & Rao, N. V. M. (2016). Modelling the Impact of Global Financial Crisis on the Indian Stock Market through GARCH Models. *Asia-Pacific Journal of Management Research and Innovation*, *12*(1), 11-22.

15. Murari, K. (2015). Exchange Rate Volatility Estimation Using GARCH Models, with Special Reference to Indian Rupee Against World Currencies. *IUP Journal of Applied Finance*, *21*(1), 22.

16. Murthy, U., Anthony, P., & Vighnesvaran, R. (2016). Factors Affecting Kuala Lumpur Composite Index (KLCI) Stock Market Return in Malaysia. *International Journal of Business and Management*, *12*(1), 122.

17. Musa, S. (2014). *An impact of exchange rate volatility on Nigeria's export: using ARDL approach* (Doctoral dissertation).

18. Oberholzer, N., & Venter, P. (2015). Univariate GARCH models applied to the JSE/FTSE stock indices. *Procedia Economics and Finance*, *24*, 491-500.

19. Olweny, T., & Omondi, K. (2011). The effect of macro-economic factors on stock return volatility in the Nairobi stock exchange, Kenya. *Economics and Finance review*, *1*(10), 34-48.

20. Sekmen, T., & Hatipoglu, M. (2016). *Financial Crises and Stock Market Behaviors in Emerging Markets*. IUP Journal of Applied Finance, 22(4), 5.

21. Srinivasan, P. (2011). Modeling and forecasting the stock market volatility of S&P 500 index using GARCH models. *IUP Journal of Behavioral Finance*, *8*(1), 51.

22. Tripathy, S., & Rahman, A. (2013). Forecasting Daily Stock Volatility Using GARCH Model: A Comparison between BSE and SSE. *IUP Journal of Applied Finance*, *19*(4), 71.